\documentclass[1p]{elsarticle} 
\usepackage{amssymb}
\usepackage{amsmath}
\usepackage{amsfonts}
\usepackage{graphicx}
\usepackage{subfigure}
\graphicspath{{images/}}

\ifx\pdfoutput\@undefined\usepackage[usenames,dvips]{color}
\else\usepackage[usenames,dvipsnames]{color}
\IfFileExists{pdfcolmk.sty}{\usepackage{pdfcolmk}}{} 
\fi

\begin{document}

\title{Rank-frequency distribution of natural languages: \\
a difference of probabilities approach}
\author[1,2]{Germinal Cocho}
\author[1,3]{R. F. Rodr\'{\i}guez\corref{cor1}}
\author[1,6]{Sergio S\'{a}nchez}
\author[1]{Jorge Flores}
\author[1]{Carlos Pineda}
\author[2,4,5]{Carlos Gershenson}
\cortext[cor1]{zepeda@fisica.unam.mx}

\address[1]{Instituto de F\'{\i}sica, Universidad Nacional Aut\'{o}noma de
M\'{e}xico, Mexico City, 01000, Mexico}
\address[2]{Centro de Ciencias de la Complejidad, Universidad Nacional Aut\'{o}noma de
M\'{e}xico,\\ Mexico City, Mexico}
\address[3]{
FENOMEC, Universidad Nacional Aut\'{o}noma de M\'{e}xico, Mexico City, Mexico
}
\address[4]{
Instituto de Investigaciones en Matem\'{a}ticas Aplicadas y Sistemas,
Universidad Nacional Aut\'{o}noma de M\'{e}xico, Mexico City, Mexico
}
\address[5]{
ITMO University, St. Petersburg, Russian Federation
}
\address[6]{
Maestr\'{\i}a en Ciencias de la Complejidad, Universidad Aut\'{o}noma de la Ciudad de M\'{e}xico,
Mexico City, Mexico
}

\begin{abstract} 
The time variation of the rank $k$ of words for six Indo-European languages is obtained using data from Google Books. For low ranks the distinct languages behave differently, maybe due to  syntaxis rules, whereas for $k>50$ the law of large numbers predominates. The dynamics of $k$ is described stochastically through a master equation governing the time evolution of its probability density, which is approximated by a Fokker-Planck equation that is solved analytically. The difference between the data and the asymptotic solution is identified with the transient, and good agreement is obtained.
\end{abstract} 

\maketitle
\section{Introduction} 

The statistical study of languages has shown an increased
interest over the last decades since the pioneering works of Zipf~\cite{zipf}
and others~\cite{mandelbrot,gell-mann,ferrer,perc}.  These
studies have focused on the rank-frequency distribution of words. Additionally, the rank
diversity distribution has recently been proposed as a novel measure to
characterize the statistical properties of languages~\cite{cocho1}. This distribution can be understood as a measure of how word ranks change in
time. This measure has also shown that the size of the language core is similar for
most languages. Within this statistical linguistic point of view, in previous
work we have introduced a simple Gaussian random walk model for the rank
diversity which reproduced some of the observed features of the evolution of
this quantity quite well~\cite{cocho1}.

Furthermore, in recent years much effort has been given to the study of
complex networks associated to physical systems, biological organisms, and
social organizations; the structure and dynamics of these networks being a matter
of intense research~\cite{albert,li-chen}. In previous works~\cite{alvarez}, we
have looked into the evolution of complex networks in terms of a master
equation (ME) describing birth-death stochastic processes along the
lines developed for ecological models~\cite{McKane,hubbel1}. We have
shown that under very general conditions in which dynamic conflict
(frustration) exists between positive and negative mechanisms, the frequency
distribution versus rank is given by the ratio of two power laws. This is
also the case for birth and death processes in ecology, or for the
excitation-inhibition process for neurons in the central nervous system.
In a large variety of systems composed by similar elements
and with similar interactions between them, the response of the system is
determined by general laws. However, there are always differences in the response of
the system in different realizations of the same experiment which can be
associated, for instance, to the large numbers law or the central limit
theorem, and follow a normal Gaussian distribution. In these cases, the
average values are the ones that depend on general laws, whereas the
differences among various realizations of the experiments obey a different
dynamics, namely, that of the great numbers law.

In this work we use this point of view to study the frequency distribution
of words in six languages~\cite{cocho1}. In particular, we analyze
the difference between the data associated with different realizations of
these conflictive dynamics and the adjustments of the real data. We do this in terms of
a time dependent probability density distribution, by assuming that the
dynamics of the rank distribution may be described by the 
ME describing an underlying one step, Markovian, birth-death stochastic
process~\cite{alvarez,jensen}. As we have shown in previous work 
\cite{supplem}, the data describing the frequency of words of several languages can be well adjusted by an asymptotic beta function. However, as it will be shown below, there is always a small difference between the data and this adjustment. Here lies the motivation of this work and one of its
main objectives is to analyze and explain the origin of this difference
within the context of the proposed stochastic model.

The outline of the present work is as follows. In Sec. 2 we define the
stochastic model and construct a ME describing the data obtained for
different Indo-European languages. Then in Sec. 3 the initial
differential-difference ME is approximated by a (nonlinear) Fokker-Planck
equation (FPE) in the continuum limit, where the discrete rank stochastic
variable may be treated as a continuous variable. Closed analytic forms for
both, the stationary and the time dependent probability density
distributions of this equation, are obtained using Pad\'{e} approximants. In
terms of these well defined approximations, we show that the analytic time-dependent solution of the FPE describes well some of the observed features.
Finally, in Sec. 4 we summarize our main results and critically discuss the
novel features and limitations of our work.
\section{Data adjustment for Indo-European languages} 

The variations of the rank $k$ in time of twenty words for three different
$k$-scales for these six languages were obtained for two centuries 
in~\cite{cocho1}; an example for English is given in Fig. 1. From the curves in~\cite{cocho1} it can be observed that the behavior of $k(t)$ is similar for all languages. Words with low rank almost do not vary in time and
as the value of $k$ increases, its variations depend on the rank itself.
Notice that there is a higher variation at all scales before year 1850. As
an example, in the case of English and for the $k$-scale between 1-30, the
variation of rank with time is very small; in contrast, for the intervals
250-1500 and 4500-15000 its variation is much larger and very irregular. This
shows that the variable $k$ exhibits different dynamics in different regions
of the $(k,t)$ space. This fact suggests that the dynamics in the
last two intervals may be described by a stochastic model for the random
variable $k$.

\begin{figure}
\begin{center}
 \includegraphics{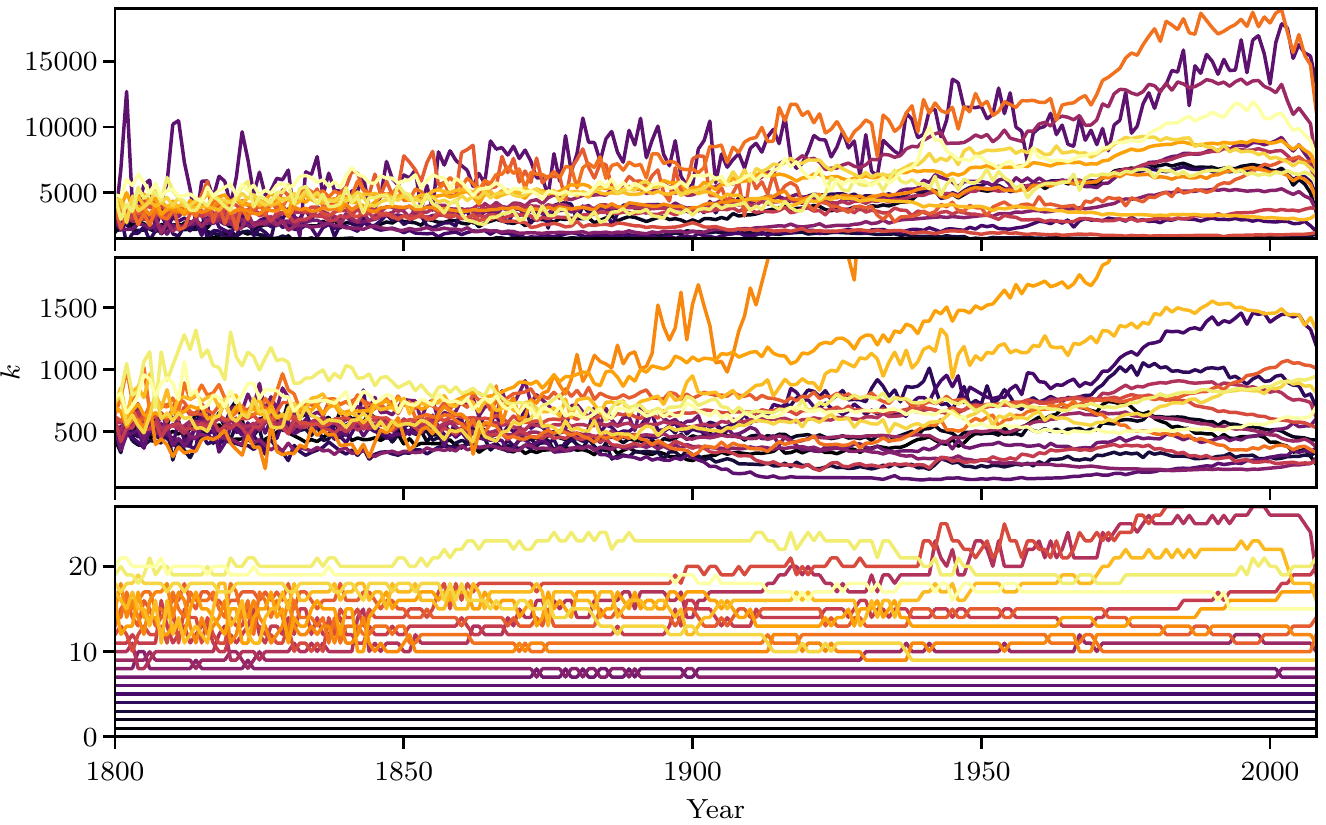}
\end{center}
\caption{Rank variation in time for twenty English words at three different
scales: In (a) for $0<k<30$, (b) $250<k<1500$, and (c) for the scale
4500-15000.}
\label{fig:<+label+>}
\end{figure}

The normalized word frequencies $f(k)$ associated with the curves in Fig. 1,
as a function of the rank $k$, were fitted with different rank distributions 
$m_{i}(k)$, $i=1,2,3,4,5$, defined by Eqs. (S1) - (S5) in~\cite{supplem}. The models $m_{i}(k)$ fit better in different
regions of $(f,k)$, but for none of them the fit is best for all languages
in all regions. However, it was found that the data adjustment is best when
the asymptotic beta function 
\begin{equation}
m_{3}(k)=\mathcal{N}_{3}\frac{\left( \overline{N}+1-k\right) ^{b}}{k^{a}}
\label{m3}
\end{equation}%
is used. Here, $a$ and $b$ are the fitting parameters, $\mathcal{N}_{3}$ is
a normalization factor and $\overline{N}$ is the total number of words. Fig.
2 shows that indeed, there is always a (small) difference between the data
and the adjustment.

\clearpage
\begin{figure}
\begin{center}
\includegraphics{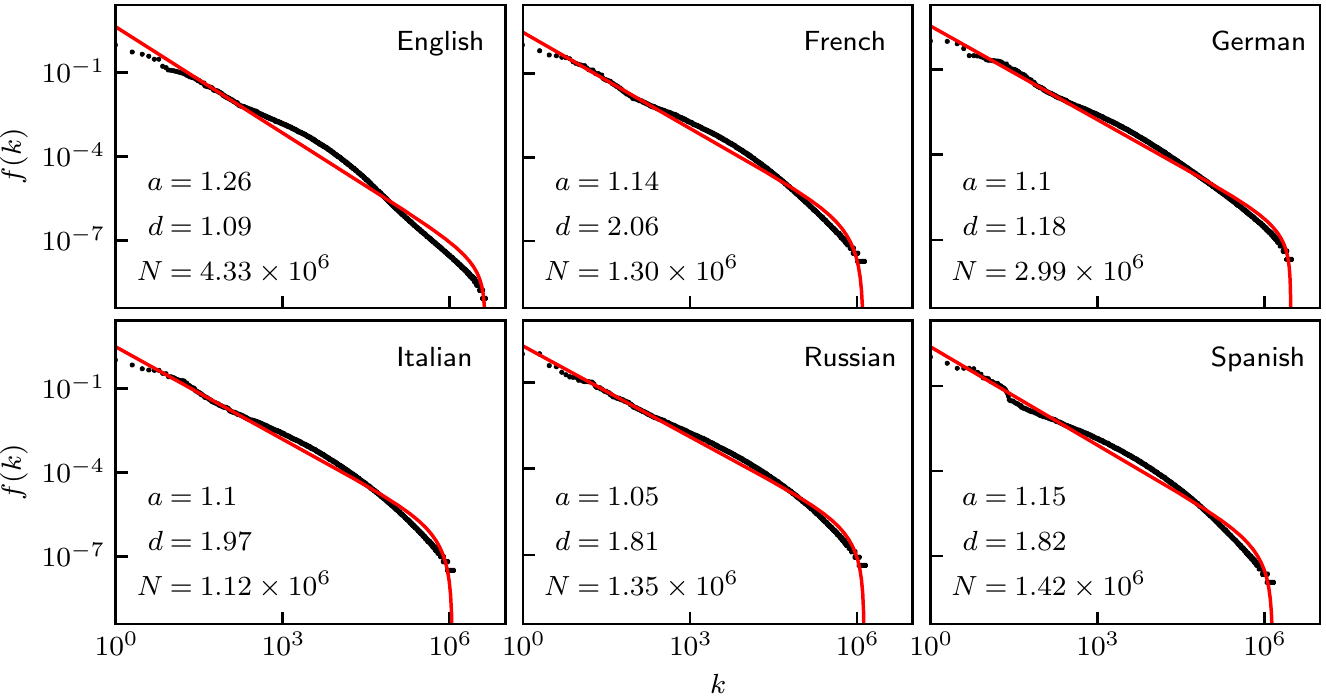}
\end{center}
\caption{
Fitting the theoretical rank distribution with the data
corresponding to the year 2008. 
}
\label{fig:<+label+>}
\end{figure}

These plots were obtained for books published in the year 2008. The
curves clearly show that none of the distributions captures satisfactorily
the entire data behavior. The usual criterion to quantify the quality of an
adjustment is to calculate the coefficient of determination (denoted by $%
R^{2}$) which is the integral of the squared difference between data and
adjustment, or the proportion of the variance in the dependent variable that
is predictable from the independent variable; if $R^{2}$ is near one, the adjustment is considered to be good. However, this quantity does
not describe which values of $k$ contribute predominantly to a specific
value of $R^{2}$. One of the objectives of the present work is to take this into
account and analyze the origin of this difference. By assuming that the
dynamics is originated by the action of multiplicative factors, in the next
section we shall describe this difference between data and adjustment in
terms of a log-normal distribution.
\section{Stochastic model} 

Given a set of words forming a text, the number of times $N(k,t)$ that a
certain word appears with the rank $k$ at time $t$ can be evaluated. If this
change in $k$ is modelled by a one-step Markovian stochastic process, and if 
$b_{k}\equiv b(k)$ and $c_{k}\equiv c(k)$ denote arbitrary functions for the
transition probabilities per unit time for the rank to increase or to
decrease in one unit, the dynamics of the probability density $P_{k}(t)\equiv $
$P(k,t)$ for the rank to have the value $k$ at time $t$ is
given by the nonlinear ME~\cite{vankampen}
\begin{equation}
\frac{\partial }{\partial t}P(k,t)=c_{k+1}P_{k+1}(t)+b_{k-1}P_{k-1}(t)-%
\left( c_{k}+b_{k}\right) P_{k}(t).  \label{1}
\end{equation}
It should be remarked that the ME is always linear in the unknown $P(k,t)$%
, and that the term nonlinear refers to the generality of the functions $%
b_{k}$ and $c_{k}$. Note that if the range of values of $k$ is finite, $%
k=0,1,2,...,N$, (\ref{1}) is meaningless for $k=0$, and this value is a
boundary of the one step process. However, by assuming

\begin{equation}
d(0)=b(-1)=0,  \label{2}
\end{equation}%
equation (\ref{1}) is still valid for $k=0$. It is convenient to rewrite (1)
in the more compact form

\begin{equation}
\frac{\partial }{\partial t}P(k,t)=\left[ \left( \widehat{E}-1\right)
d(k)+\left( \widehat{E}^{-1}-1\right) b(k)\right] P(k,t),  \label{3}
\end{equation}%
where the action of the step operators $\widehat{E}^{\pm 1}$\ over an
arbitrary function $f(k)$ is defined by

\begin{equation}
\widehat{E}^{\pm 1}f(k)=f(k\pm 1).  \label{4}
\end{equation}
\section{Fokker-Planck approximation}  

Since only in rare cases it is possible to solve the ME explicitly, we
shall assume that the changes in $k$ are small and that we are only
interested in solutions $P(k,t)$ that vary slowly with the discrete variable 
$k$. In this limit the discrete variable $k$ may be treated as a continuous
variable and the operators $\widehat{E}^{\pm 1}$\ may be replaced by a
Taylor series expansion in $k$, yielding the following nonlinear FPE
approximation for the ME~\cite{risken}

\begin{equation}
\frac{\partial P(k,t)}{\partial t}=\left\{ -\frac{\partial }{\partial k}g(k)+%
\frac{1}{2}\frac{\partial ^{2}}{\partial k^{2}}f(k)\right\} P(k,t)\equiv 
\widehat{L}(k)P(k,t).  \label{10a}
\end{equation}

Here $f(k)\equiv b(k)+c(k)$, $g(k)\equiv b(k)-c(k)$, and $\widehat{L}(k)$
defines the Fokker-Planck operator. If the dynamics takes place through
multiplicative factors, the system follows a log-normal probability
distribution. Assuming this to be the case, we write $P(k,t)$ as $P(x,t)$
with $x\equiv \log k$.

It is well known that the probability density function ($PDF$) of an
additive process depending on multiple, independent stochastic variables, is
obtained naturally through the reiterative application of convolution,
\begin{equation}
\left( P_{2}\oplus P_{1}\right) (\chi )\equiv N\int_{-\infty }^{+\infty
}\int_{-\infty }^{+\infty }d\xi _{1}d\xi _{2}P_{1}(\xi _{1})P_{2}(\xi
_{2})\delta \left( \xi _{2}+\xi _{1}-\chi \right) .  \label{11}
\end{equation}
As indicated by the $\delta$ function, the integral is performed in the
locus of an equal sum of variables. However, if we are interested in modeling a
stochastic system in which the dependence of the random variables is not
through addition but substraction, $\chi =x_{2}-x_{1}$, it can be shown that
the probabilistic outcome of $\chi $ is given by~\cite{substraction}
\begin{equation}
P(\chi )=P_{2}(\chi _{2})\oplus P_{1}(-\chi _{1}),  \label{12}
\end{equation}%
where $P_{1}(\chi _{2})$ and $P_{1}(\chi _{1})$ are the $PDF^{\prime }s$ of $%
x_{2}$ and $x_{1}$, respectively. For most non-symmetrical $PDF^{\prime }s$,
this result is sufficient to violate the validity of the central limit
theorem. However, in Ref.~\cite{substraction} it is also shown that
the new product,%
\begin{eqnarray}
\left( P_{2}\ominus P_{1}\right) (\chi ) &\equiv
&N\int_{0}^{1}\int_{0}^{1}d\xi _{1}d\xi _{2}P_{1}(\xi _{1})P_{2}(\xi
_{2})\delta \left( \xi _{2}-\xi _{1}-\chi \right)  \notag \\
&=&N\int_{\chi }^{1}d\xi P_{1}(\xi -\chi )P_{2}(\xi ),  \label{13}
\end{eqnarray}
which is the cross-correlation, describes correctly the probabilistic
outcome of $\chi$ and that the correlation function between two beta distributions is well described by a beta function. On the other hand, in \cite{cocho1} it is shown that the data of words frequency vs. rank are also well adjusted by a beta distribution. Therefore, these two observations suggest that the dynamics of frequency data as a function of rank might depend on a difference between probability distributions.

Now, according to Fig. 2 there is always a difference between the predicted
values and the data adjustments, a fact that suggests the following
analysis: If $A$ and $B$ are the probability distributions of two different
stochastic variables, $x_{1}$, $x_{2}$, and if we define
\begin{equation}
\mathcal{S}\equiv \frac{1}{2}\left( A+B\right) ,\text{ \ \ \ }\mathcal{D}%
\equiv \frac{1}{2}\left( A-B\right) ,  \label{13a}
\end{equation}%
then
\begin{equation}
A-B=\left[ \left( P_{1}+P_{2}\right) \ominus \left( P_{1}+P_{2}\right)
+\left( P_{1}-P_{2}\right) \oplus \left( P_{1}-P_{2}\right) \right] .
\label{13b}
\end{equation}

In previous works,~\cite{plos1}, we have shown that for the data associated
with the English language, the first term on the right hand side ($r.h.s.$), \emph{i.e.} the correlation, can be very well adjusted by a stationary asymptotic
distribution function $P_{\text{asym}}$ equal to the $\beta$ distribution.
Therefore, the second term on the $r.h.s.$, \emph{i.e.} the convolution, may be
identified with a Gaussian distribution. In this work we show that for the
English language data this is indeed the case. Then, as a consequence
of (\ref{13b}), $P(k,t)$ can be expressed in the general form%
\begin{equation}
P(x,t)=P_{\text{asym}}(x)+P_{1}(x,t).  \label{14}
\end{equation}

Note that for the present model $P_{\text{asym}}(x)$ may be identified with the
stationary solution of (\ref{10a}) defined by 
\begin{equation}
\widehat{L}(x)P^{st}(x)=0,  \label{15}
\end{equation}%
and that $P_{1}(x,t)$ satisfies the FPE 
\begin{equation}
\frac{\partial P_{1}(x,t)}{\partial t}=\widehat{L}(x)P_{1}(x,t).  \label{16}
\end{equation}
\subsection{Stationary solution} 

The general form of the stationary $P^{st}(x)$\ solution of (\ref{14}) is
well known~\cite{vankampen}%
\begin{equation}
P^{st}(x)=\frac{N_{0}}{f(x)}\exp \left[ 2\int_{0}^{x}\frac{g(x^{\prime })}{%
f(x^{\prime })}dx^{\prime }\right] ,  \label{16a}
\end{equation}%
where $N_{0}$\ is the integration constant which has to be chosen such that $%
P_{k}^{s}$\ is normalized. In (\ref{16a}) we restrict our calculation to the
range $1\ll k\ll N$. However, the fraction $g(x)/f(x)$ may be expressed in terms
of Pad\'{e} approximants, which are a particular type of rational fraction
approximation to the value of a function~\cite{pade}. The basic idea is to
match the Taylor series expansion as far as possible. If we denote the $L$, $%
M$ Pad\'{e} approximant to $A(x)$ by%
\begin{equation}
\left[ L/M\right] =\frac{P_{L}(x)}{Q_{M}(x)},  \label{17}
\end{equation}%
where $P_{L}(x)$ is a polynomial of degree at most $L$ and $Q_{M}(x)$\ is a
polynomial of degree at most $M$, the formal power series expansion%
\begin{equation}
A(x)=\sum_{j=0}^{\infty }a_{j}x^{j},  \label{17a}
\end{equation}%
which is unique if $\left[ L/M\right] $\ exists. The coefficients of $%
P_{L}(x)$\ and $Q_{M}(x)$\ are determined by the equations%
\begin{equation}
A(x)-\frac{P_{L}(x)}{Q_{M}(x)}=O\left( x^{L+M+1}\right) .  \label{17b}
\end{equation}%
Since we can obviously multiply the numerator and denominator by any
constant and leave $\left[ L/M\right] $ unchanged, we impose the
normalization condition%
\begin{equation}
Q_{M}(0)=1  
\end{equation}%
and write the coefficients of $P_{L}$\ and $Q_{M}$ as%
\begin{eqnarray}
P_{L}(x) &=&p_{0}+p_{1}x+...+p_{L}x^{L},  \notag \\
Q_{M}(x) &=&1+q_{1}x+...+q_{M}x^{M},  \label{17d}
\end{eqnarray}%
where $p_{0}$\ is a constant.

In Ref.~\cite{alvarez} it is shown that the fraction $g(x)/f(x)$\ may be
expressed in the form%
\begin{equation}
\frac{g_{m}(x)}{f_{r}(x)}=A_{0}+\sum_{i=1}^{N}\frac{A_{i}}{x+c_{i}},
\label{17e}
\end{equation}%
where $A_{0}$\ and $A_{i}$\ are well defined constants in terms of the
original polynomials $g(x)$ and $f(x)$,~\cite{hubbel1,plos1}, and
the stationary solution $P^{st}(x)\equiv P_{\text{asym}}(x)$ may be rewritten in
the general form%
\begin{equation}
P_{\text{asym}}(x)=\mathcal{N}\exp \left[ A_{0}x\right] \prod\limits_{i=1}^{N}%
\left( x+c_{i}\right) ^{-A_{i}},  \label{17f}
\end{equation}%
where $\mathcal{N}$\ is determined from the normalization condition and the $%
c_{i}$\ are constants determined by the above procedure.
\subsection{Time dependent solutions} 

Since the probability distribution $P_{1}(x,t)$ satisfies the FPE 
\begin{equation}
\frac{\partial P_{1}(x,t)}{\partial t}=\widehat{L}(x)P_{1}(x,t),  \label{18a}
\end{equation}%
where $\widehat{L}(x)$\ is the Fokker-Planck operator (\ref{10a}), by
defining $R(x)\equiv -g(x)$, $D(x)\equiv \frac{1}{2}f(x)$ and $%
U_{S}(x)\equiv D(x)P^{st}(x)$, (\ref{15}) can be rewritten in the more
compact form 
\begin{equation}
\frac{R(x)}{D(x)}P^{st}(x)-\frac{d}{dx}P^{st}(x)=0.  \label{18b}
\end{equation}%
If we introduce the potential $U_{1}(x)\equiv D(x)P^{st}(x)$, then 
\begin{equation}
\frac{R(x)}{D(x)}dx=\frac{dU_{1}(x)}{U_{1}(x)}  \label{18c}
\end{equation}%
and 
\begin{equation}
P^{st}(x)=\frac{1}{D(x)}\exp \left( \int \frac{R(x^{\prime })}{D(x^{\prime })%
}dx^{\prime }\right) .  \label{19}
\end{equation}%
As a result Eq. (\ref{16}) reads 
\begin{equation}
\frac{\partial P_{1}(x,\tau )}{\partial \tau }=\frac{\partial ^{2}}{\partial
x^{2}}P_{1}-\frac{\partial }{\partial x}P_{1},  \label{20}
\end{equation}%
where we have defined $\tau \equiv D(x)t$. This equation can be rewritten as
a diffusion equation by introducing the variable $V(x,t)$ through the
transformation~\cite{tijonov}%
\begin{equation}
V(x,t)\equiv D(x)P_{1}(x,t).  \label{20a}
\end{equation}

In the same way than the asymptotic solution, the function $\frac{R(x)}{D(x)}
$ in (\ref{18c}) may be also approximated by Pad\'{e} approximants, and
since the lowest order approximant $\left[ 0/0\right] $ yields a constant,
it follows that 
\begin{equation}
R(x)=\lambda _{1}G(x),\text{ \ \ \ }D(x)=\lambda _{2}G(x).  \label{20b}
\end{equation}%
Actually, we show below that this approximation is sufficient to fit the
data. Furthermore, this approximation yields 
\begin{equation}
\frac{\partial V(x,t)}{\partial \tau }=\frac{\partial ^{2}V}{\partial x^{2}}%
-K\frac{\partial V}{\partial x},  \label{20c}
\end{equation}%
where $K=\lambda _{1}/\lambda _{2}$. Finally, this equation can be reduced
to a diffusion equation by the transformation~\cite{tijonov}%
\begin{equation}
V(x,t)=U_{1}(x)\exp \left( \frac{K}{2}x+\frac{K^{2}}{4}t\right) ,
\label{20d}
\end{equation}%
which yields%
\begin{equation}
\frac{\partial }{\partial t}U_{1}(x,t)=\frac{\partial ^{2}U_{1}}{\partial
x^{2}}.  \label{21}
\end{equation}

To find the explicit analytic time dependent solution of this equation we
use the method of separation of variables and express $U_{1}(x,t)$ as%
\begin{equation}
U_{1}(x,t)=\sum_{n=1}^{\infty }A_{n}X_{n}(x)T(t).  \label{21a}
\end{equation}%
This yields the following separation equation for $T(\tau )$ 
\begin{equation}
\frac{d}{d\tau }T(\tau )=-\beta T(\tau ),  \label{23}
\end{equation}%
which for a given $T_{0}\equiv T(\tau =0)$\ has the solution%
\begin{equation}
T(\tau )=e^{-\beta \tau }T_{0},  \label{24}
\end{equation}
where $\beta $\ is an arbitrary but positive (separation) constant.
Similarly, $X(x)$ obeys the ordinary separation equation%
\begin{equation}
\frac{d^{2}}{dx^{2}}X_{n}(x)-\beta \frac{d}{dx}X_{n}(x)+d_{n}X_{n}(x)=0,
\label{25}
\end{equation}%
where the $d_{n}$ are separation constants. In terms of the variable $Y(x)$,
defined by
\begin{equation}
X(x)\equiv e^{-\beta x}Y(x),  \label{26}
\end{equation}%
the general solution of (\ref{25}) reads%
\begin{equation}
Y(x)=Ae^{-d_{1}t}\sin \left( \lambda _{0}+\sqrt{d_{1}}x\right) ,  \label{27}
\end{equation}%
where $A$\ and $\lambda _{0}$\ are, respectively, an arbitrary amplitude and
phase that have to be fixed through the initial and boundary conditions.

We now assume that its possible to replace the infinite sum (\ref{21a}) by
an effective term of the form%
\begin{equation}
\sum_{n=1}^{\infty }A_{n}X_{n}(x)T(t)\rightarrow
A_{eff}X_{eff}[x,d_{eff}(x)]e^{-d_{eff}(x)t},  \label{28}
\end{equation}%
where $X_{eff}[x,d_{eff}(x)]$ obeys the equation%
\begin{equation}
\frac{d^{2}}{dx^{2}}X_{eff}+d_{eff}(x)X_{eff}=0.  \label{29}
\end{equation}%
If we parametrize $d_{eff}(x)$ by the linear function
$d_{eff}(x)=d_{0}+d_{1}x$, the solution of (\ref{29}) is
\begin{eqnarray}
X_{eff}[x,d_{eff}(x)] &=&C_{1}\text{AiryA}_{i}\left( -\frac{d_{0}+d_{1}x}{%
d_{1}^{2/3}}\right)   \notag \\
&&+C_{2}\text{AiryB}_{i}\left( -\frac{d_{0}+d_{1}x}{d_{1}^{2/3}}\right) ,
\label{30}
\end{eqnarray}
where $\text{AiryA}_{i}$ and $\text{AiryB}_{i}$ denote the Airy functions. With these
assumptions and taking the Pad\'{e} approximant $\left[ 0/0\right] $, we may
fit the difference between the normalized word frequency $f(k)$ and the
asymptotic beta function $m_{3}(k)$ given by (\ref{m3}). For the different
languages this is shown in the plots of Fig. 3.

\begin{figure}
\begin{center}
\includegraphics{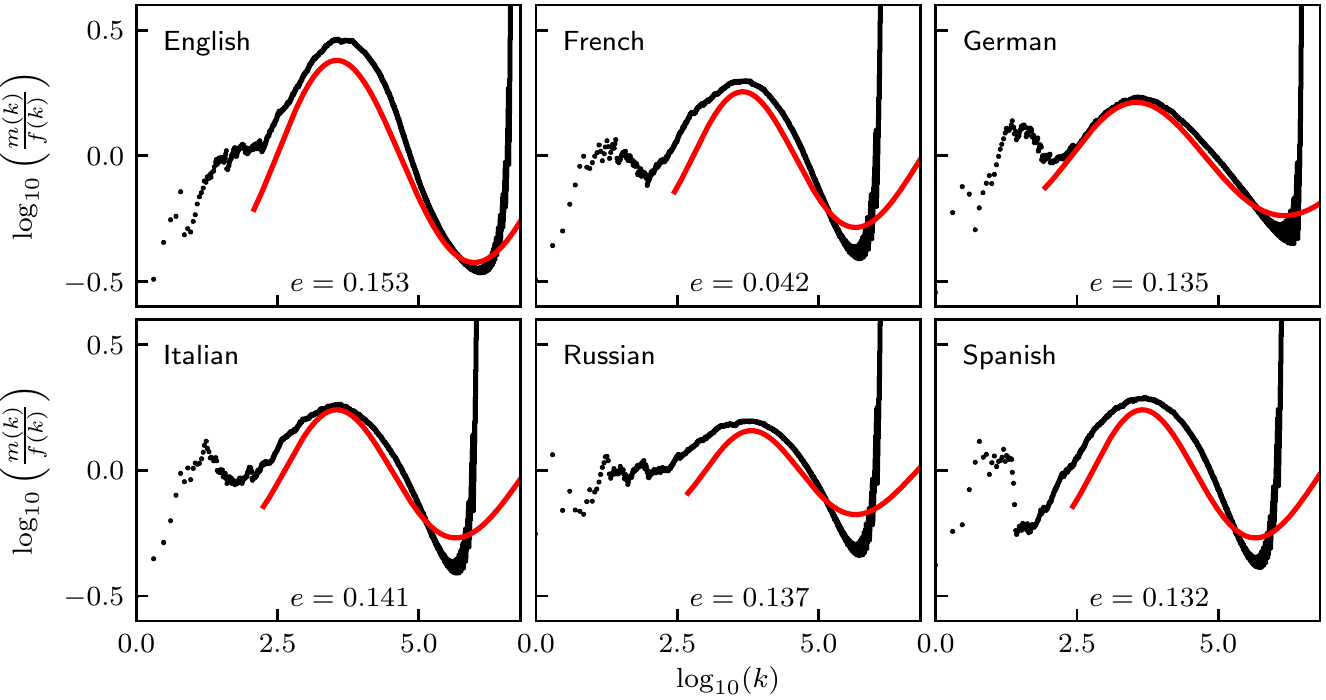}
\end{center}
\caption{The difference between the normalized word frequency $f(k)$
 and the asymptotic beta function $m_{3}(k)$, and the Airy function
between $10^{2}\leq k\leq 10^{6.5}$. The mean standard deviation, with 
respect to the fit is reported as $e$ in each plot. 
}
\label{fig:<+label+>}
\end{figure}

The curves in Fig. 3 show that for the interval $10^{2}\leq k\leq
10^{6.5}$ the dynamics of rank variation is very similar for all the
languages considered. In contrast, for $k$\emph{\ }$\leq 10^{2}$ the plots
are very different, suggesting that there are other dynamic factors that have to
be taken into account.
\section{Discussion} 

In this work we have proposed a stochastic approach to analyze the dynamics
of the rank variation ($k$) of words in time for six Indo-European
languages: English, French, German, Italian, Russian and Spanish. Based on
numerical evidence we here showed that $k$ may be regarded as a random
variable exhibiting complex dynamics in different regions of the $(k,t)$
space. This fact suggests that its dynamics could be adequately
described by a stochastic model, and we described it as a Markovian,
one-step, stochastic process arising from the conflictive dynamics of
appearance and disappearance of words. The time evolution is given by
a master equation. For the languages considered here there is
always a small difference between the data for $k$ and their adjustment. In
this work we have analyzed and proposed an explanation of the origin of this
difference
within the context of the proposed stochastic model. Actually, in previous
works we have introduced a measure of how words ranks change in time and we
have called this distribution rank diversity~\cite{plos1}. 

In this work we have used approximations to obtain stationary
and time dependent analytic solutions of the nonlinear Fokker-Planck
equation (\ref{10a}) which lead to a good fit of the data. However, there are
many open questions and further possibilities regarding a more
adequate description of the dynamics of the rank variation. It is likely
that a more complex stochastic process is able to describe other regions of (%
$k,t$) space, where the dynamics is more complex. Yet, to our knowledge
there are no other available descriptions of theoretical linguistics, and
the predicted behavior of $k$ should always comply with the analysis based
on real linguistic data. However, this remains to be assessed.

\par 
{\it Acknowledgements.--}  
Support by  projects
CONACyT 285754 and  UNAM-PAPIIT IG100518, IN-107414, and IN-107919 are acknowledged.
\end{document}